# High Energy γ-Ray Emission from the Loop I region

Jean-Marc Casandjian, Isabelle Grenier on behalf of the Fermi Large Area Telescope Collaboration
*Laboratoire AIM, CEA-IRFU/CNRS/Université Paris Diderot, Service d'Astrophysique, CEA Saclay, 91191 Gif sur Yvette, France*

Loop I is a nearby giant radio loop spanning over 100 degrees and centered on the Sco-Cen OB association. It may correspond to a superbubble formed by the joint action of stellar winds and supernova remnants. ROSAT observations revealed that the region is filled with a hot gas possibly reheated by successive supernova explosions. The brightest rim of Loop I, called the North Polar Spur (NPS), extends to the north along 30 degrees in longitude, at a distance of about 100 pc from the Sun. Early searches for high-energy γ rays associated with electrons or protons accelerated by Loop I were performed with SAS-II, COSB and EGRET. But a detector with better performance and higher statistics is required to distinguish the faint signal from the NPS from broad structures in the Galactic interstellar emission, such as the inverse Compton emission from cosmic-ray electrons scattering the interstellar radiation field. We modelled the γ-ray emission of the Galaxy and compared it to the Fermi-LAT photons detected above 300 MeV. We observe an excess of photons in the direction of Loop I. This excess exhibits a large arc-shaped structure similar to those seen in synchrotron emission.

## 1. LOOP I IN RADIO

The local interstellar medium is known to contain several bubbles or large supernova shells like the Local Bubble within which the Solar System resides. Six major Galactic loops, obtained by joining radio spurs into circles, were discovered in radio-continuum surveys. Loop I, discovered by Large, Quigley & Haslam in 1962 [1], is the most prominent one. It is a nearby giant radio loop spanning over 100 degrees and centered on the Sco-Cen OB association. The brightest feature of Loop I, called the North Polar Spur (NPS), is located in the northern hemisphere at a longitude around 30 degrees and is at a distance of approximately 100 pc from the Sun. The spur is best seen in the 408 MHz survey of Haslam *et al.* [2], in addition to the synchrotron emission along the Milky Way that arises from cosmic-ray electrons interacting with the Galactic magnetic field. From the synchrotron index map from Miville-Deschênes *et al.* [3], it appears that the Haslam map is dominated by soft electrons toward the inner plane, unlike synchrotron maps measured at higher frequency like the WMAP polarized emission at 23 GHz from Page *et al.* [4]. Since only synchrotron emission is polarized at this frequency, this map is clean of dust and free-free contribution, however the variation of the angle between the magnetic field and the line of sight directions depolarizes the signal so the map does not fully represent the total synchrotron intensity.

## 2. EARLIER SEARCHES FOR GAMMA RAYS FROM LOOP I

The search for high-energy γ rays from Loop I dates back to SAS II. In the letter to Nature called "Acceleration of cosmic rays in the Loop I 'supernova remnant' ?", Bhat *et al.* [5] claim to have detected an excess of intensity toward Loop I. More claims of detected emission were published latter with COS B and EGRET photons.

But distinguishing between the signal from Loop I and broad structures in the Galactic interstellar emission, like the inverse Compton emission from cosmic-ray electrons scattering the interstellar radiation field requires a precise Galactic diffuse emission model, and the systematic uncertainties of the models used at that time were probably much larger than the number of claimed photons. With an improved diffuse emission model, Grenier *et al.* [6] have detected a significant correlation between the EGRET photons and the 408 MHz synchrotron map toward Loop I.

If powered by supernovae, Loop I is at a very advanced stage of radiative expansion. But the case of Loop I is complex. An age of around 1 Myr was inferred from the low expansion velocity of the neutral gas surrounding the loop. However, since radio and X-ray emission are still visible, it is probable that Loop I has recently been reheated by one or more supernove.

## 3. FERMI-LAT PHOTONS AND THE GALACTIC DIFFUSE EMISSION MODEL

In order to observe extended objects with faint intensity, a precise modelling of the emission from the bulk of the Galaxy is required.

This emission is produced in the interactions of energetic cosmic-ray electrons and protons with interstellar nucleons and photons. The decay of neutral pions produced in hadron collisions, the inverse Compton scattering of the interstellar radiation field by electrons and their bremsstrahlung emission in the interstellar gas are the main contributors to the Galactic emission.

If energetic cosmic rays penetrate uniformly all gas phases, the γ-ray intensity in each direction can be modelled by a linear combination of gas column-densities (obtained from radio and infrared surveys), an inverse Compton intensity map (calculated by GALPROP [7]), and an isotropic intensity that accounts for isotropic gamma-ray background.

To account for the non-uniform cosmic-ray flux through the Galaxy, the gas column densities are distributed within six galactocentric rings. To obtain the model intensity, the emissivities for each of those components were fitted to Fermi-LAT photons.

The data used in this study were obtained in the all-sky survey mode, summing all the valid statistics of the last 14 months. We selected photons from the P6_V3_Diffuse class having a zenith angle less than $105°$ and an energy above 300 MeV so that we have both large photon counts and good spatial resolution.

Figure 1 shows both the LAT counts map with the sources removed and the model map for E>300 MeV.

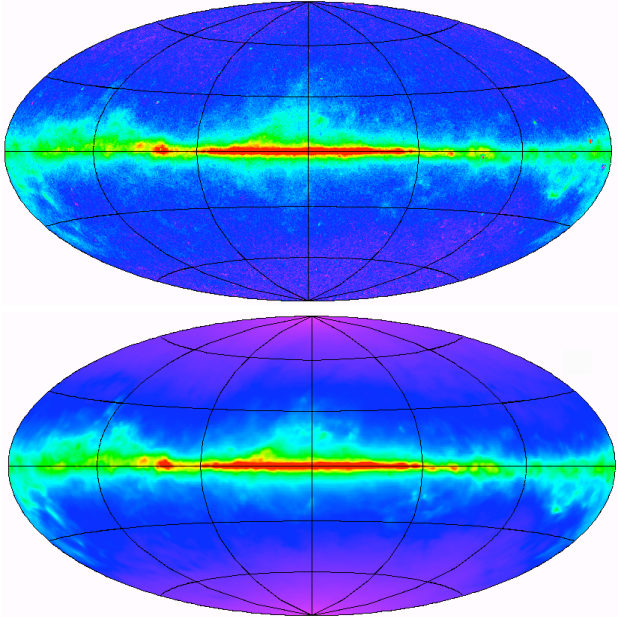

Figure 1: Fermi-LAT counts map for E>300MeV with sources removed (up) and diffuse model counts prediction for the same energy range (down).

## 4. RESULT

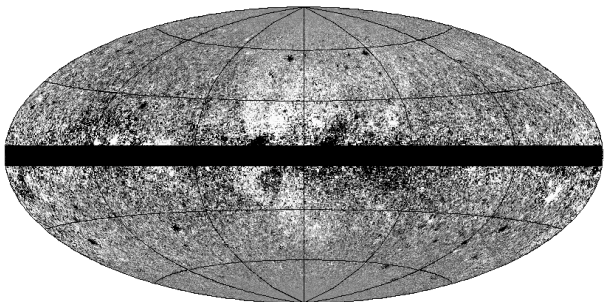

Figure 2: Preliminary residual map (LAT counts minus model) for photons with E>300 MeV.

Figure 2 shows the difference between the photon counts detected by the Fermi-LAT above 300 MeV and the counts expected from the diffuse emission model above the same energy. We have masked the Galactic plane for $|b|<5°$. We observe an excess of photons in the direction of Loop I.

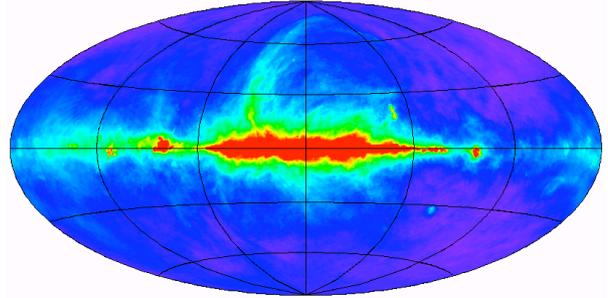

Figure 3: The Galaxy radio map at 408MHz from Haslam *et al.* [2].

As a comparison we show in Figure 3 the radio continuum at 408 MHz of Haslam *et al.* [2] predominantly produced by synchrotron emission of high energy electrons spiraling in Galactic magnetic field.

The residual map shows arc-like structures on both sides of the plane that are strongly reminiscent of the synchrotron spurs observed in both the 408MHz map and the polarized light by WMAP at 23GHz [4]. These "arcs" are best seen along the NPS (l ~ 30°) and l ~ 20° up to 50° in latitude, and at l ~ 290° from 30° to 60° in latitude. The structures seen in the residual map toward the inside of Loop I are partially coming from hydrogen gas not correctly accounted for in the model. A excess with harder spectrum possibly associated with Loop I in the northern and southern central region is also observed.

## 5. CONCLUSION

The high sensitivity of the LAT together with a precise model for the Galactic emission allowed for the first time the observation of an excess of γ rays related to Loop I. Only a careful study of the spectra and of the spatial correlation with the synchrotron data will tell us the origin of the emission. This excess is much fainter than the pion production and the bremsstrahlung contribution associated to the hydrogen gas and also fainter than the inverse Compton emission from the Milky Way as calculated by GALPROP.